# Size distribution of Au NPs generated by laser ablation of a gold target in liquid with time-delayed femtosecond pulses

E. Axente[1], M. Barberoglou[2], P.G. Kuzmin[3], E. Magoulakis[2], P. A. Loukakos[2], E. Stratakis[2], G.A. Shafeev[3], C. Fotakis[2]

[1]Lasers Department, National Institute for Lasers, Plasma, and Radiation Physics, P.O. Box MG 36, RO-77125, Bucharest-Magurele, Romania

[2]Institute of Electronic Structure and Laser, Foundation for Research & Technology—Hellas, (IESL-FORTH) P.O. Box 1527, Heraklion 711 10, Greece

[3]Wave Research Center of A.M. Prokhorov General Physics Institute of the Russian Academy of Sciences, 38, Vavilov Street, 119991 Moscow, Russian Federation

**Abstract**

The influence of delay-time between two sequential femtosecond pulses on the properties of nanoparticles obtained via laser ablation of gold in ethanol has been studied. The morphological and optical properties of the nanoparticles attained were characterized using high resolution transmission electron microscopy and UV-Vis absorption spectroscopy, respectively. Furthermore, the size distribution of nanoparticles was determined by means of a centrifugal sedimentation particle size analyzer. It is found that the time delay variations lead to corresponding changes in size distribution, plasmon resonance position as well as the rate of nanoparticles generation.

**Introduction**

Ablation of materials with intense and short laser pulses is a challenging research subject, mainly due to the reduction of the heat-affected zone when processing metals but also to the possibility of structuring wide band-gap dielectrics by nonlinear absorption. Moreover, irradiation of metals with short laser pulses has shown to be an efficient tool to produce particles with sizes in the range of a few nanometers [1]. Nanoparticles (NPs) can be additionally produced by short pulse laser ablation of a solid target in liquid media [REFs]. This method gives a unique opportunity to solve the cytotoxic effects reported for the NPs produced by chemical synthesis [2].

The fundamental mechanisms involved in short pulse laser ablation are not yet fully understood and depend on several parameters, including laser wavelength, pulse duration, and laser fluence[3]. Although many theoretical [4] and experimental [5] studies have been devoted to the interpretation of the physical phenomena occurring during the intense short-pulse laser interaction with materials, some processes like the dissipation of absorbed energy into the lattice and corresponding material removal mechanisms, are still under investigations. Recently, the double-pulse technique, consisting in the application of two time-delayed sequential laser pulses, has been used in order to get a better insight into the temporal characteristics of the different ablation mechanisms [6-12].

Compared to material's ablation in vacuum or in low pressure backgrounds, laser ablation in liquids has attracted significant attention during last several years mainly because it allows producing "pure" NPs of a large variety of compounds comprising neither surface-active substances nor counter-ions [14 - 15]. Various laser sources can be used for NPs generation using this technique, from IR to UV, while their pulse duration spans from tens of nanoseconds to tens of femtoseconds [13]. During laser ablation in liquids the laser pulse passes through the liquid layer that should be transparent for the laser wavelength and melts the surface of the target. The thin liquid layer adjacent to the melted surface is heated up to high temperature and expands over the melt providing its ejection into the surrounding liquid. As a result, the dispersed melted material remains in the liquid forming thus a colloidal solution of NPs. Upon ablation of metallic targets, the laser radiation is absorbed by free electrons, and the subsequent surface melting occurs after the thermalization of the electrons within the lattice. In case of metal's ablation with fs laser pulses, the time of electron-phonon relaxation is on the order of the pulse duration itself, and thus, both melting and expansion of the melt into the surrounding liquid occurs when the laser pulse is already over.

If the ablation of a metal target is carried out with two consecutive time-delayed fs pulses, short delay times, comparable with the time of electron-phonon relaxation, could result in the production of higher concentration of free electrons that will simultaneously transfer their energy to the lattice. For longer delays one might expect that the resulting NPs have the same properties as those produced by two independent laser pulses. In other words, short delays between fs pulses should influence the electronic sub-system while the result of ablation at long delays would coincide with that of ablation with individual pulses.

The aim of this study is to demonstrate for the first time the influence of the delay time between two consecutive fs pulses on the synthesis and properties of NPs during laser ablation of an Au target in ethanol medium.

**Experiment**

The ablation experiments were carried out with a Ti:Sapphire laser source (FemtoPower, CompactPro) delivering pulses of 200 fs duration at a repetition rate of 1 kHz and 800 nm wavelength. The Gaussian laser beam was focused, through a liquid layer, onto the target surface with a lens of 300 mm focal length. A Michelson interferometer was used for the generation of the two time-delayed pulses; the laser beam was splitted in two parts, each beam following different paths and then reflected on two mirrors at 90° incidence. One of the mirrors was placed on a motorized, computer controlled stage. By adjusting this mirror's position the length of one beam path is changed, and thus the delay time ($t_d$) between the two laser pulses can be varied accordingly. As a result, at the interferometer output, two time-delayed laser pulses of equal energy were obtained. In the experiments presented here $t_d$ was ranged from 0 to 10 ps.

A gold target of 99.9% purity was placed on the bottom of a Pyrex cell and covered with a 1-2 mm layer of absolute ethanol without addition of any surface-active substances. The cell was fixed onto a rotating stage to avoid craters formation onto the target surface and to ensure identical irradiation conditions for the subsequent pulses during laser exposure. The exposure time and the liquid volume were kept identical in all irradiation experiments. Following ablation experiments, the optical properties of Au NPs in ethanol colloidal solutions were characterized using UV-Vis Absorption Spectroscopy (Perkin Elmer, Lambda 950). The size distribution of Au NPs in colloid was measured with the aid of a centrifugal sedimentation particle size analyzer (CPS). The morphology of Au NPs was studied using High Resolution Transmission Electron Microscopy (HR TEM, JEOL). For this purpose droplet of the different colloid solutions were placed on carbon-coated TEM grids.

**Results**

A typical HR TEM image of Au NPs generated in our experimental conditions is shown in Fig. 1. As can be observed, NPs are spherical and polycrystalline, comprising nanocrystals exhibiting several crystallographic orientations.

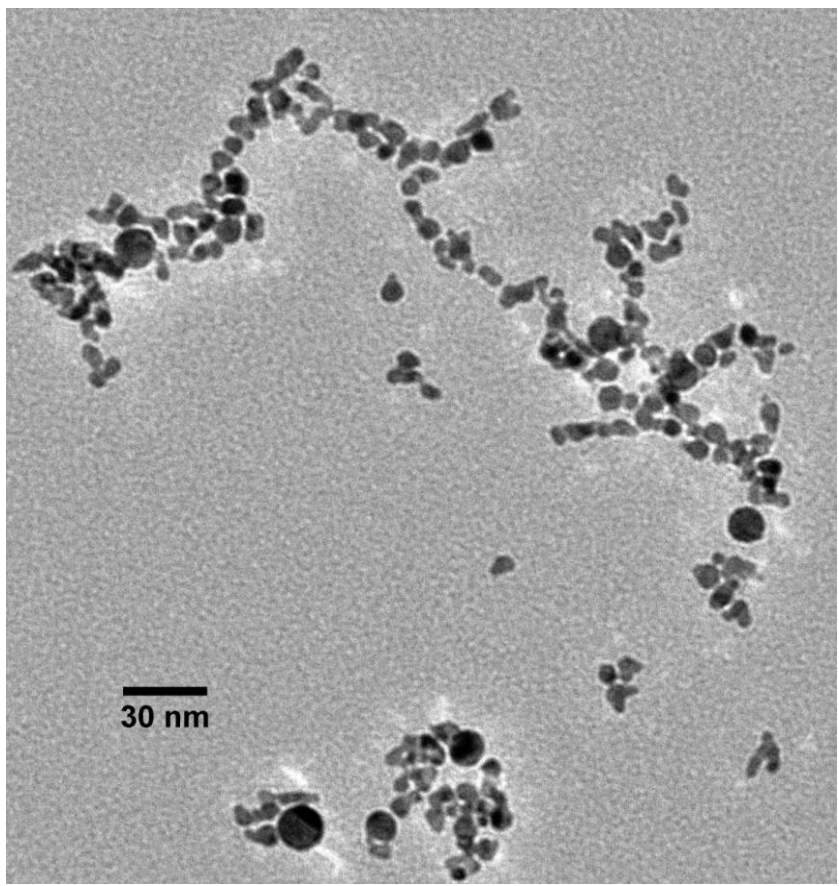

a

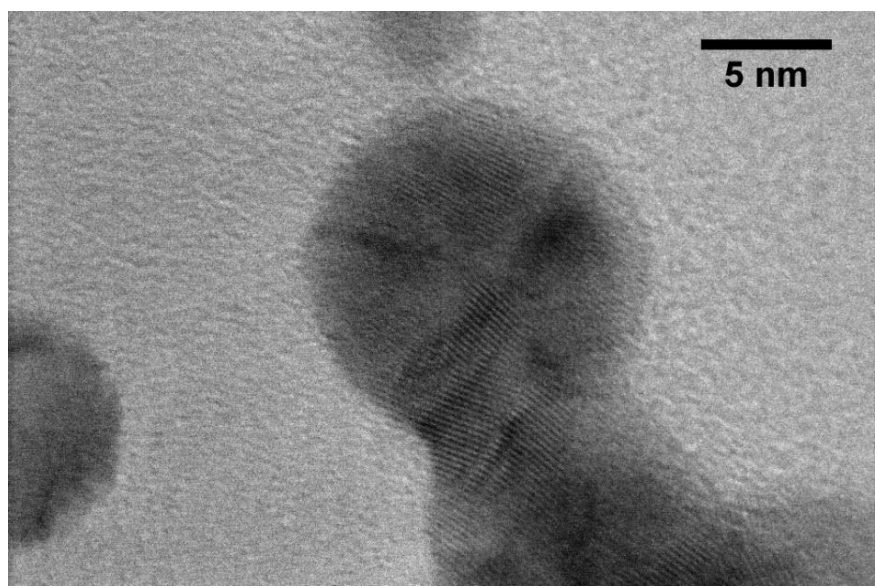

b

*Fig. 1. Typical HR TEM image of Au NPs generated by laser ablation of a gold target in ethanol. General view (a), single Au nanoparticle (b). Crystallographic planes with different orientations are clearly visible. Delay between the two fs pulses is 10 ps.*

The extinction spectra of colloidal solutions (Fig. 2) indicate a plasmon resonance absorption peak at 520 - 540 nm In agreement to theoretical predictions [16].

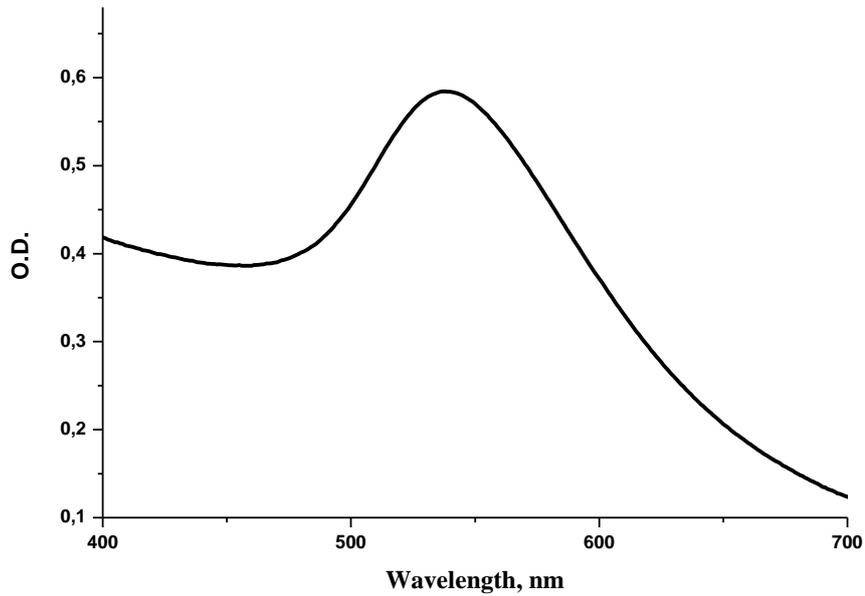

*Fig. 2. Optical density of colloidal solution of Au NPs in ethanol in the vicinity of the plasmon resonance generated with pulse delay of 200 fs.*

As shown in Fig. 3 the position of the plasmonic peak is systematically shifted to the red region, and the magnitude of this shift becomes highest for the shortest used corresponding to 100 fs.

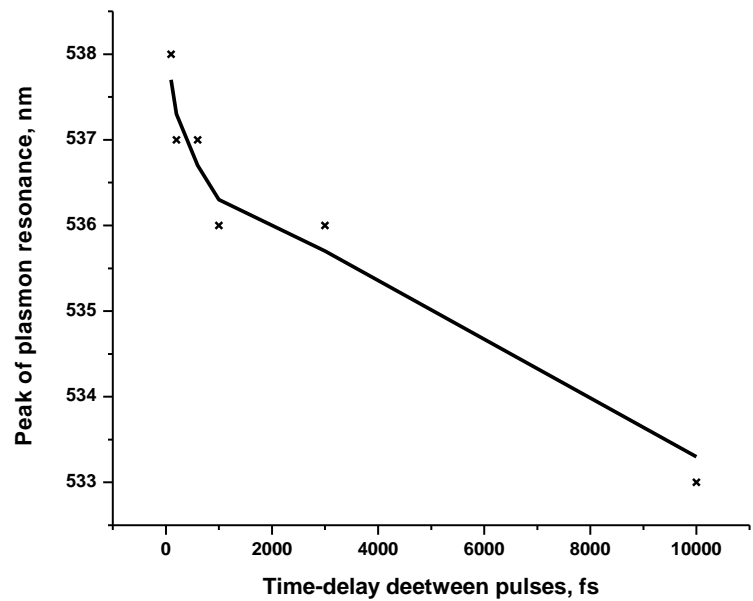

*Fig. 3. Dependence of the maximum of plasmon resonance of Au NPs on the time-delay between fs laser pulses.*

Usually the red shift of the plasmon resonance of spherical Au NPs is attributed to the increase of the average size of NPs. This denotes that short delays between pulses lead to the generation of relatively larger NPs. Such evidence was confirmed by the size distribution analysis performed on the respective NPs colloids, shown Fig. 4.

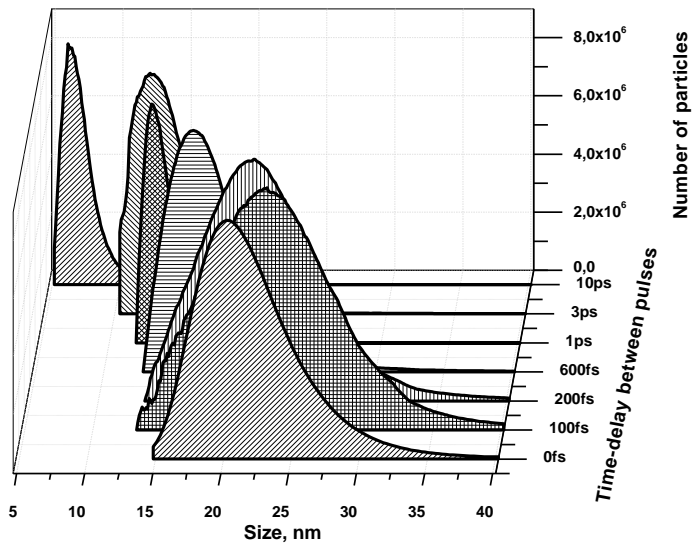

*Fig. 4. The family of normalized size distribution functions of Au NPs versus the time-delays between fs laser pulses.*

Indeed, Fig. 5 depicts that upon increase of the $t_d$ between the two pulses the position of the size distribution maximum decreases (Fig. 5) and reaches a plateau value around 8 nm that corresponds to the maximum of the average size of Au NPs generated by each individual single fs pulse of the same fluence.

It is pertinent to note that the exposure of the target is accompanied by generation of so called white continuum in the liquid. No systematic studies of this continuum were done in the frame of this work. However, qualitatively, the intensity of the white continuum varies with the delay between the two pulses. Its intensity is highest at short delays of order of 100-200 fs, with further increase of delays the intensity of the white continuum radiation becomes negligible.

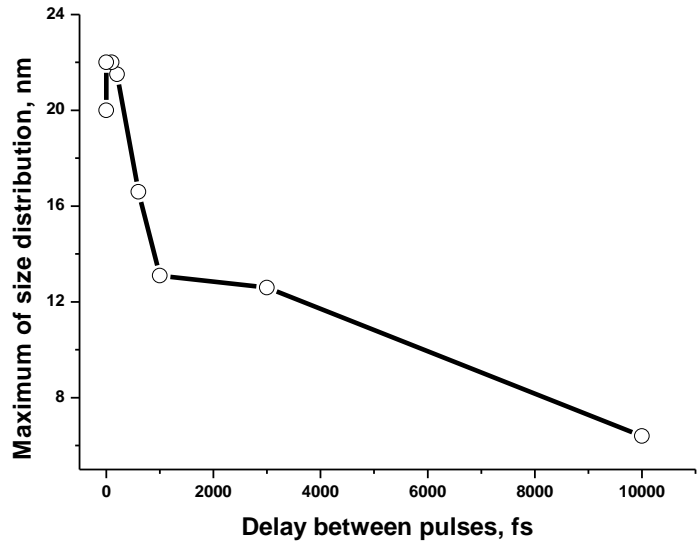

*Fig. 5. Dependence of the maximum size of Au NPs on the time-delays between fs laser pulses.*

The areas under the size distribution curves shown in Fig. 4 represent the efficiency of NPs generation for the different delay times at the fixed irradiation time used. The corresponding dependence of the formation efficiency versus $t_d$ is presented in Fig. 6 showing that the rate of NPs generation is low at small delay times between pulses and then increases linearly up to 2 ps where it reaches a plateau value. It should be noted that the efficiency of NPs generation exceeds the one gained with a single pulse of the half-fluence but less than by a factor of 2.

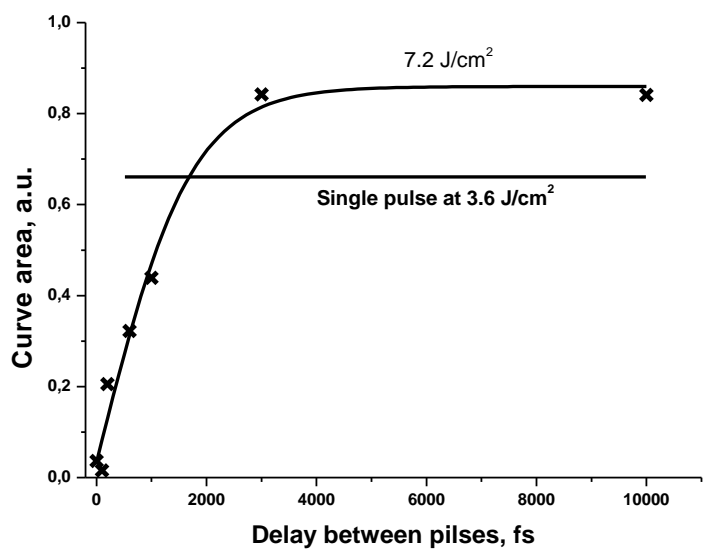

*Fig. 6. Relative efficiency of NPs generation as a function of the delay between the two fs laser pulses.*

**Discussion**

An important observation presented in Fig. 5 is the existence of an initial increase of NPs size at very low delay times, followed by a significant drop when $t_d$ reaches several hundreds of fs. In a simplified assumption, our results can be interpreted using the two-temperature model (TTM), as soon as material's properties depend on both electron, $T_e$, and lattice, $T_l$, temperatures. In case of laser ablation of metals, the energy is first absorbed by the free electrons within a layer determined by the optical penetration depth. Subsequently, the electrons diffuse into the bulk before the electron-phonon coupling takes place. Thus, for short delays, i.e. 100-300 fs, electrons become hot while the lattice remains cold and an inter-band relaxation of electrons excited by absorbed laser photons occurs. The arrival of the second pulse generates additional electrons that can diffuse further into the target. As a consequence, the concentration of electrons generated by two subsequent pulses at short time-delays is higher than that generated by a single pulse with the same peak power. This results in the increase of penetration depth of electrons inside the target before transferring their energy to the lattice, which in turn leads to a higher thickness of the molten layer and thus generation of larger NPs. For time-delays exceeding 600 fs each laser pulse can be considered as independent. As proposed by Kanavin *et al.*, the electron heat conductivity $k_e$ decreases with the lattice temperature, i.e. $k_e \propto T_e/T_l$, for moderate electron temperatures [17]. Thus, upon increasing $t_d$, $T_l$ decreases and $k_e$ decreases and the melt thickness becomes lower, which in turn, leads to the generation of smaller NPs.

Theoretical simulations in the framework of the TTM are on-going and preliminary results obtained are in accordance to the above experimental observations. Our results indicate that the double-pulse technique is a promising tool to control the size and shape of NPs produced by laser ablation of materials in liquid media.

**Conclusion**

Ablation of Au target in ethanol using two time-delayed consecutive femtosecond pulses lead to generation of colloidal solution of NPs, the properties of which vary with the time-delay used. Variation of $t_d$ leads to corresponding changes in size distribution, plasmon resonance peak position and rate of NPs generation. The experimental observations were qualitatively explained in the framework of the TTM, while theoretical simulations are on-going for a deeper understanding of the different processes taking place. Our work shows that laser ablation of materials in liquid media using sequential pulses is a promising technique to control the size and shape of the colloidal NPs produced.

**Reference:**

1. B. Liu, Z. Hu, and Y. Chen, Appl. Phys. Lett. **90**, 044103 (2007).
2. A. V. Kabashin and M. Meunier, Journal of Photochemistry and Photobiology A: Chemistry **182,** 330 (2006).
3. D. Bäuerle, Laser Processing and Chemistry. – Berlin : Springer, 3rd ed., 2000.
4. S. I. Anisimov and B. S. Luk'yanchuk, Physics -Uspekhi **45**, (3) 293 (2002).
5. A.-C. Tien, S. Backus, H. Kapteyn, M. Murnane, and G. Mourou, Phys. Rev. Lett. **82**, 3883 (1999).
6. S. Noël and J. Hermann, Appl. Phys. Lett. **94**, 053120 (2009).
7. A. Semerok and C. Dutouquet, Thin Solid Films **453–454**, 501 (2004).
8. D. Scuderi, O. Albert, D. Moreau, P. Pronko, and J. Etchepare, Appl. Phys. Lett. **86**, 071502 (2005).
9. S. Amoruso, R. Bruzzese, X. Wang, and J. Xia, App. Phys. Lett. **93**, 191504 (2008).
10. T. Donnelly, J. G. Lunney, S. Amoruso, R. Bruzzese, X. Wang, and X. Ni, J. Appl. Phys. **106**, 013304 (2009).
11. M. Spyridaki, E. Koudoumas, P. Tzanetakis, C. Fotakis, R. Stoian, A. Rosenfeld, and I. Hertel, Appl. Phys. Lett. **83**, 1474 (2003).
12. Z. Hu, S. Singha, Y. Liu, and R. J. Gordon, Appl. Phys. Lett. **90**, 131910 (2007).
13. G.A. Shafeev, Formation of nanoparticles under laser ablation of solids in liquids, in: *Nanoparticles:New Research*, editor Simone Luca Lombardi, 2008, pp. 1-37, Nova Science Publishers, Inc.
14. S. Besner, A.V. Kabashin, F.M. Winnik, and M. Meunier, Appl. Phys. A **93**, 955 (2008).
15. Emmanuel Stratakis, Marios Barberoglou, Costas Fotakis, Guillaume Viau, Cecile Garcia, and Georgy A. Shafeev, Opt. Express **17**, 12650 (2009).
16. J. A. Creighton, D. G. Eadon, J. Chem. Soc. Faraday Trans., **87**(24), 3881-3891 (1991).
17. A. Kanavin, I. Smetanin, V. Isakov, Y. Afanasiev, B. Chichkov, B. Wellegehausen, S. Nolte, C. Momma, and A. Tünnermann, Phys. Rev. B **57**, 14698 (1998).